# Revised Modularity Index to Measure Modularity of OSS Projects with Case Study of Freemind

Andi Wahju Rahardjo Emanuel
Informatics Bachelor Program
Maranatha Christian University
Bandung, Indonesia

Daniel Jahja Surjawan
Informatics Bachelor Program
Maranatha Christian University
Bandung, Indonesia

## ABSTRACT
Open Source Software (OSS) Projects are gaining popularity worldwide. Studies by many researchers show that the important key success factor is modularity of the source code. This paper presents the revised Modularity Index which is a software metrics to measure the modularity level of a java-based OSS Projects. To show its effectiveness in analyzing OSS Project, the Modularity Index and its supporting software metrics are then used to analyze the evolution of Freemind mind mapping OSS Project. The analysis using Modularity Index and its supporting metrics shows the strength and weaknesses of the Freemind OSS Projects.

## 1. INTRODUCTION
Open Source Software (OSS) Projects are gaining popularity worldwide. The previously considered bazaar-like and experimental software building effort has developed into one of the mainstream software development methodology challenging the already established software engineering methodology. There are already many examples of successful OSS Projects to date such as Linux Operating System, Mozilla Browser, Apache Web Server, etc. which are comparable or better than the proprietary counterparts. The successes of these projects are studied and one of the main key success factors is modularity of the source codes [1].

Modularity is considered one of the key success factors of OSS Projects since its high correlation with quality [2]. The highly modular OSS Projects are also considered to have high quality, and in turn the high quality OSS Projects also depends on the sustainable communities [3]. The first author has previously proposed the first quantitative measure of modularity called Modularity Index [4]. In this paper, the revision of the Modularity Index is proposed and the revised measure is used to analyze the evolution of Freemind OSS Project to demonstrate its effectiveness in detecting the strength and weaknesses of the OSS Project.

This paper is structured into four main sections. The first section describes the current research relating to software modularity, modularity in OSS Projects, and the challenge in measuring software. The second section describes the detail formulation of the revised Modularity Index. The new formulation is used to measure and analyze the evolution of Freemind in the third section. The last section contains conclusion and acknowledgement.

## 2. RELATED WORKS
The success and failure of OSS Projects have been the subject of many studies and the software modularity is believed to be the major cause of success in these OSS Projects [1]. In order for a particular OSS Project to be successful, the project must allow a new developer to develop a new module [5], and using four possible approaches [6]. The modularity of successful OSS Projects are believed to be better than their proprietary counterparts [7].

Measuring software and specifically its modularity level presents significant challenge. The first successful attempts to measure object oriented metrics are Chidamber and Kemerer [8]. Other proposed methods are the effort estimation [9], and coupling-based structural metrics [10]. In measuring modularity, analytical approach could be used [11]. The integrated measure of modularity in OSS Projects is the Modularity Index [4]. After some observations and considerations, authors are revising the previous formulation which will enhance better comprehension of this software metrics.

## 3. REVISION OF MODULARITY INDEX
Similar to the previous formulation of Modularity Index, the formulation of the revised Modularity Index should begin from class level, then progresses into the Package (module) level and finally into system level. The Modularity Index is intended as the single number to determine the level of modularity in the whole system level.

### 3.1 Class Level Modularity
There are three main components that are responsible for class level modularity which are LOC Quality ($LOC_Q$), Function Quality ($F_Q$), and Cohesion Quality ($H_Q$). The first two components ($LOC_Q$ and $F_Q$) are already introduced previously and the third component ($H_Q$) is a new measure.

The $LOC_Q$ is a normalized value that determines the quality of a class based on the number of Non-Commenting Lines of Code (NCLOC) in the class. This value is based on the observation of selected 50 java-based OSS Projects [4] and the formulation is shown here for completeness purpose:

$$LOC_Q = 0.0138\, NCLOC + 0.310 \; for\; NCLOC \leq 50 \dots (1)$$

$$LOC_Q = \frac{1}{(NCLOC - 50)^{1.969}} \; for\; NCLOC > 50 \dots (2)$$

Where:
    $LOC_Q$ = LOC Quality
    NCLOC = Non-Commenting Lines of Code

Similarly, the $F_Q$ is a normalized value that determines the quality of a class based on the number of Functions (methods) in the class. This value is also based on the observation of selected 50 java-based OSS Projects [4].

$$F_Q = 0.1836\,F + 0.0820 \quad for\ F \leq 5 \quad \ldots(3)$$

$$F_Q = \frac{1}{(F - 4.83)^{2.691}} \quad for\ F > 5 \quad \ldots(4)$$

Where:
  $F_Q$ = Function Quality
  $F$ = Function / method

The third component is Cohesion Quality ($H_Q$) which is the normalized value that determines the quality of a class based on its cohesion value (LCOM4). This Cohesion Quality replaces the value LCOM4 which is directly used in the class Quality formulation [4]. By using the selected 50 java-based OSS Project and using the inverse polynomial square fit, the formulation of Cohesion Quality is shown in formula 5.

$$H_Q = \frac{1}{LCOM4^{2.216}} \quad \ldots(5)$$

Where:
  $H_Q$ = Cohesion Quality
  LCOM4 = Lack of Cohesion Metrics 4

Finally, the formulation of $c_Q$ is composed of these three components with the weight of each components are differentiated based on the fact that NCLOC and F are found to have high correlation, and the LCOM4 is independent since it has low correlation with NCLOC and F.

$$c_Q = 0.25\,LOC_Q + 0.25\,F_Q + 0.5\,H_Q \quad \ldots(6)$$

Where:
  $c_Q$ = class Quality
  $LOC_Q$ = LOC Quality
  $F_Q$ = Function Quality
  $H_Q$ = Cohesion Quality

The class Quality is a normalized value that determines the quality of a particular class. The maximum value of class Quality is achieved when a class has 50 NCLOC, 5 Functions (methods) and perfect cohesion (LCOM4 equals to 1).

Table 1 shows the summary of differences between the previous and the revised version of class level modularity.

**Table 1. Comparison in class level**

| Parameter | Previous Version | Revised Version |
|---|---|---|
| $c_Q$ | unchanged | |
| $F_Q$ | unchanged | |
| $H_Q$ | not used | introduced to replace the LCOM4 |
| $c_Q$ | Composed of $LOC_Q$, $F_Q$ and LCOM4 | Composed of $LOC_Q$, $F_Q$ and $H_Q$ as shown in equation 6 |

## 3.2 Package Level Modularity
There is no difference for formulation of Package Quality. The Package Quality is the average of class quality in that package:

$$P_Q = \frac{\sum_{i=1}^{j} c_{Qi}}{\sum_{i=1}^{j} c_i} \quad \ldots(7)$$

Where:
  $P_Q$ = Package Quality
  $c_{Qi}$ = i-th class Quality
  $c_i$ = i-th class

Even though there is no observed similarity of the number of classes in each Package, it can be observed from the 50 selected OSS Projects that most OSS Projects has the 10 to 16 classes per Package.

## 3.3 System Level Modularity
Similar to the previous publication [4], the components for system level modularity are System Architecture ($S_A$) and Package Quality ($P_Q$). The value of $S_A$ determine the quality of the architecture of the system, this value is determined by two parameters which are Package Coupling and Package Cohesion. Package Coupling (Cij) is defined as the count of dependency from classes in one package to the classes in other packages in the system, whereas the Package Cohesion (Cii) is defined as the count of dependency from classes in one package to classes in that packages including dependencies into the class itself. The formulation of the $S_A$ is based on the formulation of similar value but using entropy measures [12].

$$S_A = \sqrt{\frac{\sum_{i=1}^{d} c_{ii}^2}{\sum_{i=1}^{d}\sum_{j=1}^{d} c_{ij}^2}} \quad \ldots(8)$$

Where:
  $S_A$ = System Architecture
  $C_{ii}$ = Package Cohesion
  $C_{ij}$ = Package Coupling (if $i \neq j$)

The value of $S_A$ is a normalized value from 0 to 1. The 0 value means that the system has none of Package Cohesion, where as the maximum value of $S_A$ is achieved in a system with have many of Package Cohesion and none of Package Coupling.

## 3.4 Formulation of Modularity Index
The Modularity Index defined as the product of System Architecture and the Average of Package Quality.

$$M_I = S_A \frac{\sum_{i=1}^{j} P_{Qi}}{\sum_{i=1}^{j} P} \quad \ldots(9)$$

Where:
  $M_I$ = Modularity Index
  $S_A$ = System Architecture
  $P_{Qi}$ = i-th Package Quality
  $P_i$ = i-th Package

The Modularity Index is a normalized value with possible value from 0 to 1. The maximum value of $M_I$ is reached when $S_A$ is one and average $P_Q$ is also one. The average $P_Q$ of one is achieved when all of the Package Quality in the system are also one.

This formulation of Modularity Index is different from the previous formulation in which the value of $M_I$ was always increasing as the number of Packages increases [4]. This new formulation should increase the comprehensability of the software metrics and also makes it possible to compare the Modularity Indexes of java-based OSS Project with different sizes (classes and Packages).

Table 2 shows the summary of differences between the previous and the revised version of system level modularity.

**Table 2. Comparison in class level**

| Parameter | Previous Version | Revised Version |
|---|---|---|
| Package Coupling | unchanged | |
| Package Cohesion | unchanged | |
| $S_A$ | unchanged | |
| $M_I$ | Always increasing value (not divided by number of Packages) | Normalized value as shown in equation 9 |

## 4. CASE STUDY: FREEMIND

To demonstrate the effectiveness of the revised Modularity Index, this measure and its components are then used to analyze the evolution process of a java-based OSS Projects that have high number of downloads in sourceforge.net portal which is Freemind mind mapping software. The analysis is based on the measurement of software metrics indicating modularity which are NCLOC, Packages, classes, average class per Package, Functions, average Function per class, average NCLOC per class, $S_A$, average $P_Q$, and finally $M_I$.

Freemind is a Java-based OSS Project used to draw mind mapping. Mind mapping, which is an alternative way to perform brainstorming by using graphical representation in the paper or digital document, in the modern era is popularized by Tony Buzan [13]. The latest version of Freemind is version 0.9 with many of unstable versions before the milestone version 1.0.0 which is not yet reached. The study of the evolution of Freemind should give some insight about the evolution process before an OSS Project reach version 1.0.0 which should be the project's first significant milestone.

The data collection process is using SONAR (http://www.sonarsource.org) which is an application suite to collect many software metrics in Java-based Projects. In order for a particular version of a project to be analyzed by SONAR, the source code should be able to be compiled using ANT in which the measurement using SONAR called within build.xml build script. There are 22 versions that are collected and compiled starting from version 0.4 dated 7 July 2001 until version 1.0.0 Beta 8 dated 7 October 2012. Table 3 shows the detail of the Freemind versions that are collected and measured.

**Table 3. Lisf of Freemind Projects**

| No | Projects | Versions | Released Date |
|---|---|---|---|
| 1 | Freemind | 0.4 | 7 July 2001 |
| 2 | Freemind | 0.5 | 24 August 2002 |
| 3 | Freemind | 0.6 | 1 February 2003 |
| 4 | Freemind | 0.6.1 | 8 February 2003 |
| 5 | Freemind | 0.6.5 | 4 September 2003 |
| 6 | Freemind | 0.6.7 | 25 October 2003 |
| 7 | Freemind | 0.7.1 | 21 March 2005 |
| 8 | Freemind | 0.8.0 | 7 September 2005 |
| 9 | Freemind | 0.8.1 | 26 February 2008 |
| 10 | Freemind | 0.9 | 18 February 2011 |
| 11 | Freemind | 1.0.0 Alpha 1 | 26 March 2011 |
| 12 | Freemind | 1.0.0 Alpha 3 | 14 April 2011 |
| 13 | Freemind | 1.0.0 Alpha 4 | 21 May 2011 |
| 14 | Freemind | 1.0.0 Alpha 5 | 26 June 2011 |
| 15 | Freemind | 1.0.0 Alpha 6 | 30 September 2011 |
| 16 | Freemind | 1.0.0 Alpha 7 | 8 November 2011 |
| 17 | Freemind | 1.0.0 Alpha 8 | 18 December 2011 |
| 18 | Freemind | 1.0.0 Beta 1 | 17 February 2012 |
| 19 | Freemind | 1.0.0 Beta 2 | 29 April 2012 |
| 20 | Freemind | 1.0.0 Beta 3 | 9 May 2012 |
| 21 | Freemind | 1.0.0 Beta 5 | 10 June 2012 |
| 22 | Freemind | 1.0.0 Beta 7 | 5 September 2012 |
| 23 | Freemind | 1.0.0 Beta 8 | 7 October 2012 |

### 4.1 Evolution of NCLOC

The mostly used software metrics showing the size of a project is Non-Commenting Lines of Codes (NCLOC). Figure 1 shows the evolution of NCLOC of 23 versions of Freemind.

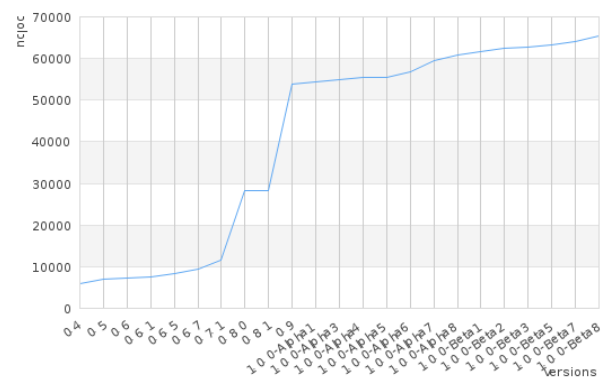

**Fig 1: Evolution of NCLOC in Freemind**

Above figure shows that the NCLOC of Freemind has grown 11-fold from 5835 in version 0.4 to 65,408 in version 1.0.0 Beta 8. The growth of NCLOC from version 0.8.0 to 0.8.1 is the same (other metrics also shows the identical value) showing the vacuum period of the project for about 2.5 years. The growth of NCLOC in the beta versions of Freemind stabilized nearing 65K showing the increasing maturity of the source code.

## 4.2 Evolution of Packages

Modularity Index defines a 'module' in java-based OSS Projects is a Package. Figure 2 shows the evolution of Packages in Freemind

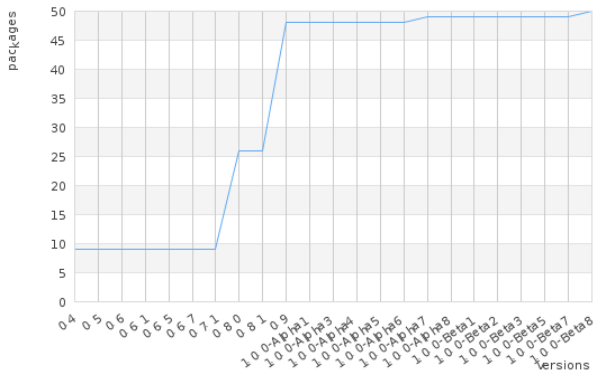

**Fig 2: Evolution of Packages in Freemind**

The number of Packages has grown from 9 packages in version 0.4 to 50 packages in version 1.0.0 Beta 8 indicating the significant feature improvement of the project. The growth of Packages in the beta versions is showing stability in the number of modules indicating the stability in the number of features in the project.

## 4.3 Evolution of classes

Figure 3 shows the evolution of classes in Freemind OSS Project.

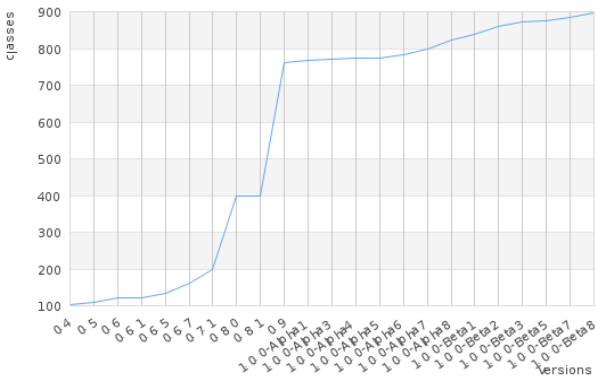

**Fig 3: Evolution of classes in Freemind**

The number of classes has grown from 104 in version 0.4 to 898 in version 1.0.0 Beta 8. The number of classes is also showing stability point in the beta versions of Freemind indicating the stability of the source codes.

## 4.4 Evolution of average classes per Package

Figure 4 shows the evolution of the average class per package in Freemind.

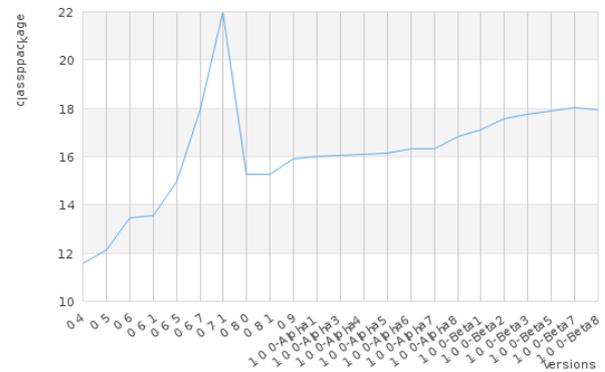

**Fig 4: Evolution of average classes per Package in Freemind**

The average classes per package have shown an interesting result. The value had risen significantly from version 0.4 to version 0.7.1. Starting from version 0.8.0 onward, the number of classes per package has rises moderately until it reach a stable value at 18 classes per package in 1.0.0 beta versions. This number is slightly higher than from the common values from our prevous observations [4] of between 10 – 16 classes per package.

## 4.5 Evolution of average Function per class

Figure 5 shows the evolution of average Function (method) per class in Freemind.

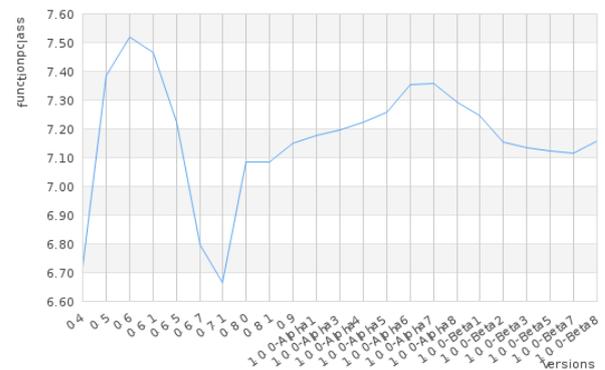

**Fig 5: Evolution of Function per class in Freemind**

Above figure shows that the average Function per class varies as the version number increases. In version 0.8.0 onward, the average number of Function per class has stabilizes at about 7 Functions per class. The optimal value of Function per class based on our previous observation [4] should be about 5, so that the value of 7 Functions per class is higher which indicated un-optimized coding practices.

## 4.6 Evolution of NCLOC per class

Figure 6 shows the evolution of average NCLOC per class in Freemind

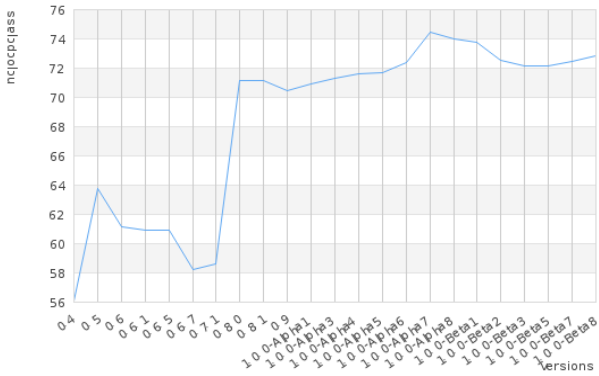

**Fig 6: Evolution average NCLOC per class in Freemind**

Above figure shows that average NCLOC per class in Freemind tends to increase as the version number increases. The average NCLOC per class is still in acceptable values which are about 80 NCLOC per class.

## 4.7 Evolution of average $P_Q$

Figure 7 shows the evolution of average Package Quality in Freemind.

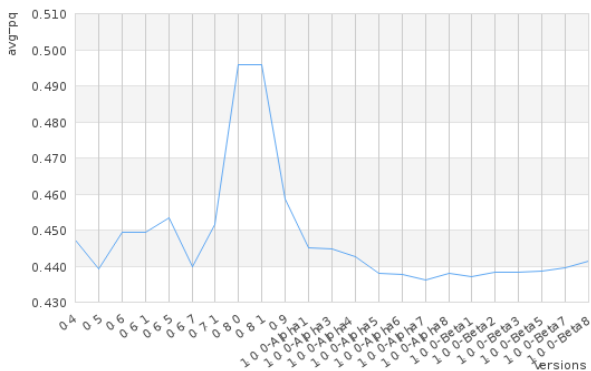

**Fig 7: Evolution of average Package Quality in Freemind**

Figure above shows that the average Package Quality in many versions of Freemind are relatively stable at about 0.4 with some exceptions are shown in version 0.8.0 and 0.8.1 in which the values are nearing 0.5. This value is low since the maximum number of average Package Quality should be nearing 1. This may be due to the number of Function per class that is nearing 7 even though the number of NCLOC per class is already in optimal value of about 80.

## 4.8 Evolution of $S_A$

Figure 8 shows the evolution of System Architecture value in Freemind.

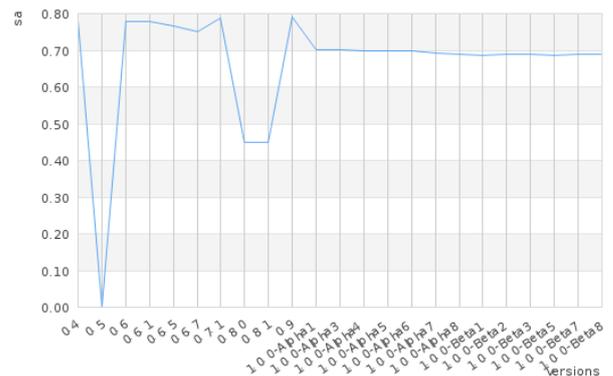

**Fig 8: Evolution of System Architecture in Freemind**

It is shown the value of $S_A$ in Freemind are stabilizing at value 0.7 at the latest versions of Freemind. This value is a high value of system architecture showing the dependency structure of the project already implementing the "maximize cohesion and minimize coupling" principle.

## 4.9 Evolution of $M_I$

Finally, figure 9 shows the evolution of Modularity Index in Freemind.

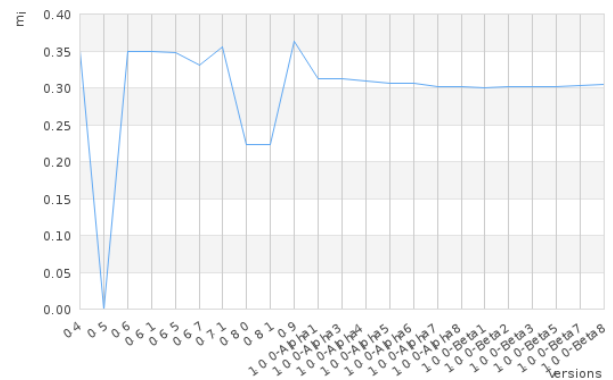

**Fig 9: Evolution of Modularity Index in Freemind**

Figure 9 above shows that the value of Modularity Index in latest versions of Freemind stabilizes at value 0.3. Since Modularity Index is the product of $S_A$ and average $P_Q$, the low number of Modularity Index is mostly due to the low number of average $P_Q$ even though the number of $S_A$ is high.

## 4.10 Analysis of Evolution in Freemind

By analyzing the value of Modularity Index and all supporting software metrics, the evolution of Freemind OSS projects has some strength and weaknesses that can be observed. The strength is especially shown in the system architecture in which the values of SA especially in the Beta versions are about 0.7 which is a high value. The Freemind projects already applying "maximize cohesion and minimize coupling" principles. The other strength are the average value of NCLOC per class is nearing ideal value which is about 72 – 74 NCLOC per class.

The weaknesses of the Freemind projects are mainly shown in the coding practices. The average value of classes per Package is about 18 classes per Package which is considered too high compared to the standard value of 10 – 16 classes per package. The number of Functions (methods) per class of about 7 is also too high compared to the standard value of 5 Functions per class.

## 5. CONCLUSIONS

In this paper, the revised Modularity Index is presented, which is a composite software metrics to measure the modularity level of a java-based OSS Projects. There are several modifications from the previous formulation, such as the introduction of Cohesion Quality ($H_Q$) instead of using LCOM4 cohesion metrics directly, and also the new formulation of Modularity Index which is the product of System Architecture ($S_A$) and average Package Quality ($P_Q$). The new formulation makes the Modularity Index value is a normalized value with possible value between 0 (no modularity) to 1 (perfect modularity).

To show the effectiveness of the revised Modularity Index, the evolution of Freemind is analyzed using this software metrics and the supporting metrics. It can be shown that the Freemind OSS Project has a high value of System Architecture ($S_A$) but a low number of average Package Quality ($P_Q$) which causes a low number of Modularity Index. The improvement that should be made to the project such as the reduction of the average Function per class and the reduction of the number of classes per Package while maintaining the high level of System Architecture.

Our future research should include the further analysis of evolutions in other java-based OSS Projects using Modularity Index. The application of Modularity Index in other object-oriented software project such as using C sharp, C++, and OOP PHP are also possible.

## 6. ACKNOWLEDGMENTS

Authors would like to thank Maranatha Christian University (http://www.maranatha.edu) that provides funding for the research.